\documentclass[12pt,a4paper,twoside]{article}
\newenvironment{changemargin}[2]{%
\begin{list}{}{%
\setlength{\leftmargin}{#1}%
\setlength{\rightmargin}{#2}%
}%
\item[]}
{\end{list}}
\usepackage{graphicx}
\usepackage{color}
\usepackage{lscape}
\usepackage{hyperref}
\usepackage{epstopdf}
\usepackage{lscape}
\usepackage{amsmath}
\usepackage{hyperref}
\usepackage{amsmath}
\usepackage{setspace}
\begin{document}
\baselineskip=0.30in
{\bf \LARGE

\begin{changemargin}{-1.2cm}{0.5cm}
\begin{center}
{Shifted Tietz-Wei oscillator for simulating the atomic interaction in diatomic molecules}
\end{center}

\end{changemargin}
}
\vspace{4mm}
\begin{center}
{\Large{\bf Babatunde J. Falaye $^a$$^{,}$$^\dag$$^{,}$}}\footnote{\scriptsize E-mail:~ fbjames11@physicist.net;~ babatunde.falaye@fulafia.edu.ng\\ \dag{Corresponding} author}\Large{\bf ,} {\Large{\bf Sameer M. Ikhdair $^b$$^{,}$}}\footnote{\scriptsize E-mail:~ sameer.ikhdair@najah.edu;~ sikhdair@gmail.com.} \Large{\bf and} {\Large{\bf Majid Hamzavi $^c$$^{,}$}}\footnote{\scriptsize E-mail:~ majid.hamzavi@gmail.com }
\end{center}
{\small
\begin{center}
{\it $^\textbf{a}$Applied Theoretical Physics Division, Department of Physics, Federal University Lafia,  P. M. B. 146, Lafia, Nigeria.}
{\it $^\textbf{b}$Department of Physics, Faculty of Science, an-Najah National University, New campus, P. O. Box 7, Nablus, West Bank, Palestine.}
{\it $^\textbf{c}$Department of Physics, University of Zanjan, Zanjan, Iran.}
\end{center}}

\begin{abstract}
\noindent
The shifted Tietz-Wei (sTW) oscillator is as good as traditional Morse potential in simulating the atomic interaction in diatomic molecules. By using the Pekeris-type approximation to deal with the centrifugal term, we obtain the bound-state solutions of the radial Schr\"odinger equation with this typical molecular model via the exact quantization rule (EQR). The energy spectrum for a set of diatomic molecules ($NO \left(a^4\Pi_i\right)$, $NO \left(B^2\Pi_r\right)$, $NO \left(L'^2\phi\right)$, $NO \left(b^4\Sigma^{-}\right)$, $ICl\left(X^1\Sigma_g^{+}\right)$, $ICl\left(A^3\Pi_1\right)$ and $ICl\left(A'^3\Pi_2\right)$ for arbitrary values of $n$ and $\ell$ quantum numbers are obtained. For the sake of completeness, we study the corresponding wavefunctions using the formula method.
\end{abstract}


{\bf PACs No.}: 03.65.Fd, 03.65.Ge, 03.65.Ca, 03.65-W

\section{Introduction}
\label{sec1}
By employing the dissociation energy and the equilibrium bond length for a diatomic molecule as explicit parameters, Jia et al \cite{J1} generated improved expressions for some well-known potentials including Rosen-Morse, Manning-Rosen, Tietz and Frost-Musulin potential energy functions. These authors found that the well-known Tietz potential function is conventionally defined in terms of five parameters but it actually has only four independent parameters. Furthermore, the Wei \cite{G1} and Tietz potential functions \cite{G2} are exactly same solvable empirical functions.

Wang et al \cite{J2} also generated improved expressions for two versions of the Schi\"oberg potential energy function which are the Rosen-Morse and Manning-Rosen potential functions. By choosing the experimental values of the dissociation energy, equilibrium bond length and equilibrium harmonic vibrational frequency as inputs, the authors obtained the average deviations of the energies calculated with the potential model from the experimental data for five diatomic molecules, and find that no one of six three-parameter empirical potential energy functions is superior to the other potentials in fitting experimental data for all molecules examined.

\begin{figure}[!htb]
\centering\includegraphics[height=100mm,width=180mm]{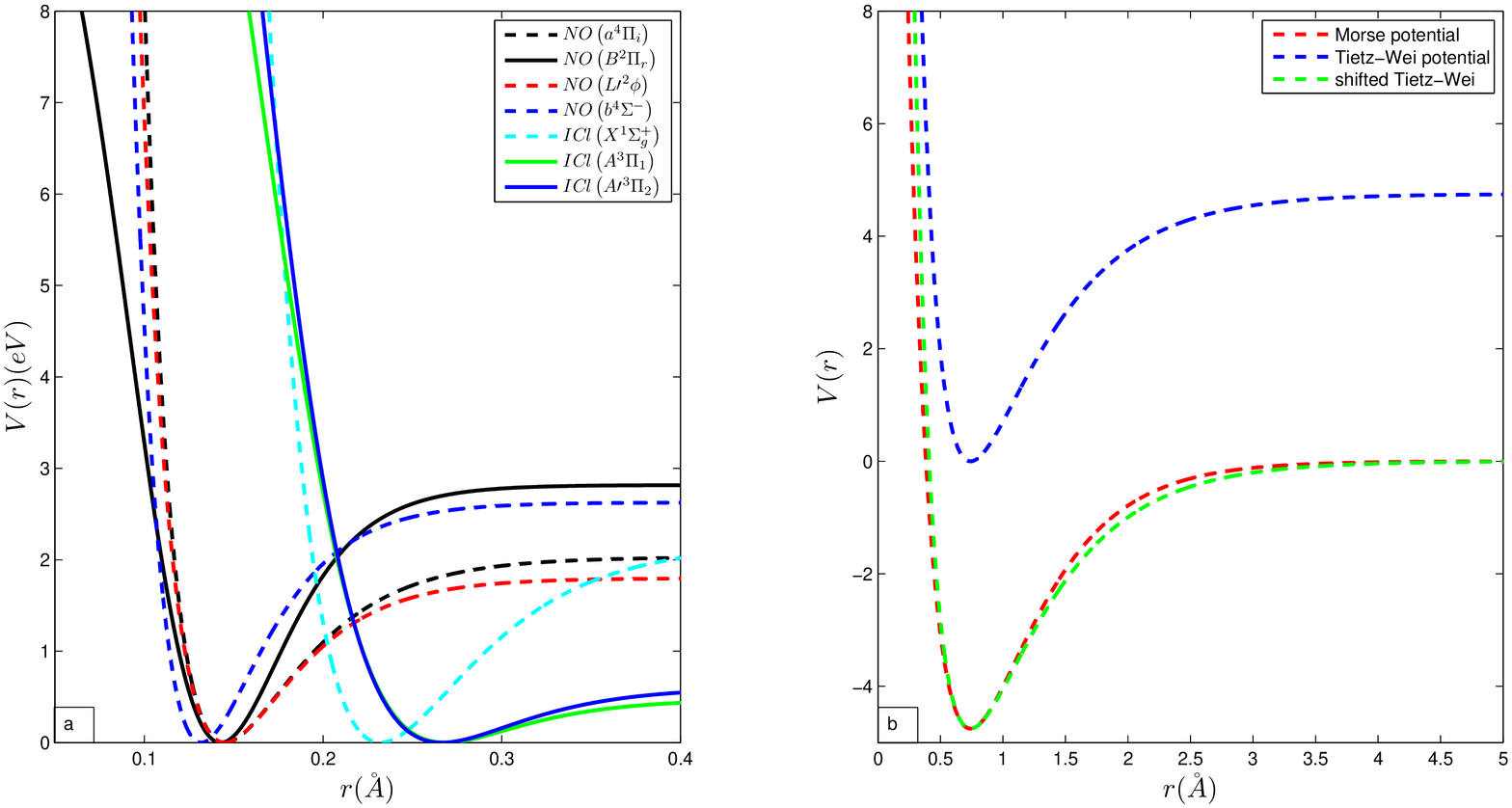}
\caption{{\protect\footnotesize (a) Shape of Tietz-Wei diatomic molecular potential for different diatomic molecules. (b) Shape of the sTW, TW and Morse oscillator potentials for $H_2 \left(X^1\Sigma^+_g\right)$ diatomic molecule with $c_h = 0.170066$, $b_h = 1.61890\AA^{-1}$, $r_e = 0.7416\AA$ and $V_e = 38318cm^{-1}$.}}
\label{fig1}
\end{figure}

All these efforts were made in an attempt to find a most suitable molecular potential in its description of diatomic molecules. Following Refs. \cite{J1,J2}, we suggest sTW as a modification for the TW [2-8]. This potential can be written as
\begin{equation}
V(r)=V_e\left[\frac{2(c_h-1)e^{-b_h(r-r_e)}-(c_h^2-1)e^{-2b_h(r-r_e)}}{\left(1-c_he^{-b_h(r-r_e)}\right)^2}\right],
\label{E1}
\end{equation}
where $b_h=\beta(1-c_h)$, $r_e$ is the molecular bond length, $\beta$ is the Morse constant, $V_e$ is the potential well depth and $c_h$ is an optimization parameter obtained from ab initio or Rydberg-Klein-Rees (RKR) intramolecular potentials. $r$ is the internuclear distance. When the potential constant approaches zero, i.e. $c_h\rightarrow 0$, the sTW potential reduces to the Morse potential \cite{J4}. This  potential is just the TW potential shifted by dissociation energy $D_e$. The shape of this potential is shown in figure \ref{fig1}a for different molecules. 

Figure \ref{fig1}b compare between TW diatomic molecular potential, sTW diatomic molecular potential and the Morse potential using the parameters set for $H_2 \left(X^1\Sigma^+_g\right)$ diatomic molecule. As it can be seen from this plot, the shifted Tietz-Wei and the Morse potentials are very close to each other for large values of $r$ in the regions $r\approx r_{e}$ and $r>r_{e}$, but they are very different at $r\approx 0$. This implies that the shifted Tietz-Wei potential is as good as traditional Morse potential and better than the Tietz-Wei potential in stimulating the atomic interaction for diatomic molecules.

The scheme of our presentation is as follows. In the next section we give basic ingredient of exact quantization rule and all necessary formulas for our calculations. We solve the radial Schr\"odinger equation for the sTW and also obtain the rotational-vibrational energy spectrum for some diatomic molecules in section 3. Finally, results and conclusions are presented in section 4.

\section{A Brief Review of the Exact Quantization Rule}
Here we give a brief review on the EQR. The details can be found in refs. [10-24].  In 2005, Ma and Xu \cite{BJ3, BJ4} by carefully studying one-dimensional Schr\"odinger equation, have extended results to three-dimensional case by simply making the replacements $x\rightarrow r$ and $V(x)\rightarrow V_{eff}(r)$:
\begin{equation}
\int_{r_a}^{r_b}k(r)dr=N\pi+\int_{r_a}^{r_b}\phi(r)\left[\frac{dk(r)}{dr}\right]\left[\frac{d\phi(r)}{dr}\right]^{-1}dr,\ \ \ \ k(r)=\sqrt{\frac{2\mu}{\hbar^2}[E-V_{eff}(r)]}, 
\label{E2}
\end{equation}
where $r_A$ and $r_B$ are two turning points determined by $E=V_{eff}(r)$. The $N=n+1$ is the number of the nodes of $\phi(r)$ in the region $E_{n\ell}=V_{eff}(r)$ and is larger by one than the number $n$ of the nodes of wave function $\psi(r)$.  The first term $N\pi$ is the contribution from the nodes of the logarithmic derivative of wave function, and the
second is called the quantum correction.

In this approach, the energy spectrum equation is obtained by solving the two integrals involved in equation (\ref{E2}). This quantization rule has been used in many physical systems to obtain the  exact solutions of many exactly solvable quantum systems [10-24]. EQR is a very important foundation to proper quantization rule (PQR) \cite{DO1}.
\section{The Energy Spectrum}
To study any quantum physical model characterized by the diatomic molecular potential given by equation (\ref{E1}), we need to solve the following Schr\"{o}dinger equation for spherically symmetric potential in any arbitrary dimensional space:
\begin{equation}
\left(\frac{d^2}{dr^2}+\frac{D-1}{r}\frac{d}{dr}-\frac{\ell(\ell+D-2)}{r^2}+\frac{2\mu}{\hbar^2}(E_{n\ell}-V(r))\right)\psi_{n,\ell,m}(r,\Omega_D)=0.
\label{E3}
\end{equation}
Now, by defining the wavefunction $\psi_{n,\ell,m}(r,\Omega_D)$ as $r^{(1-D)/2}R_{n\ell}(r)Y_{\ell m}(\theta,\phi)$ and taking $V(r)$ as the sTW diatomic molecular potential, the radial part of equation (\ref{E3}) can be found as
\begin{equation}
\frac{d^2R_{n\ell}(r)}{dr^2}+\frac{2\mu}{\hbar^2}\left[E_{n\ell}-V_e\left[\frac{2(c_h-1)e^{-b_h(r-r_e)}-(c_h^2-1)e^{-2b_h(r-r_e)}}{1-c_he^{-b_h(r-r_e)}}\right]^2-\frac{\left(\tilde{\eta}^2-\frac{1}{4}\right)\hbar^2}{2\mu r^2}\right]R_{n\ell}(r)=0,
\label{E4}
\end{equation}
where $n$, $\ell$ and $E_{n\ell}$ denote the principal quantum numbers, orbital angular momentum numbers  and the bound state energy eigenvalues of the system under consideration (i.e., $E_{n\ell}<0$ ),  respectively. The parameter $\eta=\ell+\frac{1}{2}(D-2)$ which is a linear combination of the spatial dimensions $D$ and the angular momentum quantum number $\ell$. It is well known that for $\ell=0$, problem in the form (\ref{E4}) is exactly solvable.  But for $\ell\neq0$, it is not due to the centrifugal barrier. Therefore, in order to solve the above equation for $\ell\neq0$ states, Hamzavi et al \cite{J12} found that the following formula is a good approximation scheme to deal with the centrifugal barrier:
\begin{equation}
\frac{1}{r^2}\approx\frac{1}{r_e^2}\left(D_0+D_1\frac{e^{-b_h(r-r_e)}}{1-c_he^{-b_h(r-r_e)}}+D_2\frac{e^{-2b_h(r-r_e)}}{\left(1-c_he^{-b_h(r-r_e)}\right)^2}\right),
\label{E5}
\end{equation}
with
\begin{subequations}
\begin{eqnarray}
D_0&=&1-\frac{1}{\alpha}(1-c_h)(3+c_h)+\frac{3}{\alpha^2}(1-c_h)^2, \ \ \ \ \ \ \lim_{c_h\rightarrow 0} D_0=1-\frac{3}{\alpha}+\frac{3}{\alpha^2}\\
D_1&=&\frac{2}{\alpha}(1-c_h)^2(2+c_h)-\frac{6}{\alpha^2}(1-c_h)^3, \ \ \ \ \ \ \lim_{c_h\rightarrow 0} D_1=\frac{4}{\alpha}-\frac{6}{\alpha^2}\\
D_2&=&-\frac{1}{\alpha}(1-c_h)^3(1+c_h)+\frac{3}{\alpha^2}(1-c_h)^4, \ \ \ \ \ \ \lim_{c_h\rightarrow 0} D_2=-\frac{1}{\alpha}+\frac{3}{\alpha^2},
\label{E6}
\end{eqnarray}
\end{subequations}
Constant $\alpha=b_hr_e$ has been introduced for mathematical simplicity. Now, by inserting this approximation into equation (\ref{E4}) and then introducing a new transformation of the form $r\rightarrow\zeta=\frac{r-r_e}{r_e}$ through the mapping function $\zeta=f(r)$ with $r$ in the domain $\left[\left.0, \infty\right.\right)$ or $\zeta$ in the domain $[-1, \infty]$, we obtain the following second order differential equation:
\begin{eqnarray}
&&\frac{1}{r_e^2}\frac{d^2R_{n\ell}(\zeta)}{d\zeta^2}+\frac{2\mu}{\hbar^2}\left[E_{n\ell}-V_{eff}(\zeta)\right]R_{n\ell}(\zeta)=0,\ \ \ \ \ \ \mbox{with}\label{E7}\\
&&V_{eff}(\zeta)=\left[\mathcal{A}+\frac{\mathcal{B}}{e^{\alpha\zeta}-c_h}+\frac{\frac{\ell(\ell+1)D_2}{r_e^2}+\frac{2\mu V_e}{\hbar^2}(c_h-1)^2}{\left(e^{\alpha\zeta}-c_h\right)}+\mathcal{F}{\left(e^{\alpha\zeta}-c_h\right)^2}\right],\nonumber\\
&&\mathcal{A}=\left(\frac{\tilde{\eta}^2-\frac{1}{4}}{r_e^2}\right)\frac{D_0\hbar^2}{2\mu},\ \ \mathcal{B}=\left(\frac{\tilde{\eta}^2-\frac{1}{4}}{r_e^2}\right)\frac{D_1\hbar^2}{2\mu}+2V_e(c_h-1),\nonumber\\
&&\mathcal{F}=\left(\frac{\tilde{\eta}^2-\frac{1}{4}}{r_e^2}\right)\frac{D_2\hbar^2}{2\mu}+V_e(c_h-1)^2.\nonumber
\end{eqnarray}
If we define $\tilde{\varsigma} =\frac{1}{e^{\alpha\zeta}-c_h}$, then we can obtain the two turning points $\tilde{\varsigma}_a$ and $\tilde{\varsigma}_b$ as well as their sum and product properties by solving $V_{eff}(\zeta)-E_{n\ell}=0$ or $V_{eff}(\tilde{\varsigma})-E_{n\ell}=0$ as:
\begin{eqnarray}
&&\tilde{\varsigma}_a=-\frac{\mathcal{B}}{2\mathcal{F}}-\frac{1}{2\mathcal{F}}\sqrt{\mathcal{B}^2-4\mathcal{F}(\mathcal{A}-E_{n\ell})},\ \ \ \mbox{and}\ \ \ \tilde{\varsigma}_b=\frac{\mathcal{B}}{2\mathcal{F}}+\frac{1}{2\mathcal{F}}\sqrt{\mathcal{B}^2-4\mathcal{F}(\mathcal{A}-E_{n\ell})}\label{E8}\\
&&\tilde{\varsigma}_a+\tilde{\varsigma}_b=-\frac{\mathcal{B}}{\mathcal{F}},\ \ \ \ \tilde{\varsigma}_a\tilde{\varsigma}_b=\frac{\mathcal{A}-E_{n\ell}}{\mathcal{F}}  \ \ \ \mbox{and} \ \ \  k(y)=\sqrt{\frac{2\mu \mathcal{F}}{\hbar^2}}\left[-\left(\tilde{\varsigma}-\tilde{\varsigma}_a\right)\left(\tilde{\varsigma}-\tilde{\varsigma}_b\right)\right]^{1/2}\nonumber
\end{eqnarray}
 Now, we can write the non-linear Riccati equation for the ground state is as
\begin{equation}
-\alpha\frac{\tilde{\varsigma}(1+\tilde{\varsigma})}{r_e}\phi_0'(\tilde{\varsigma})+\phi_0^2(\tilde{\varsigma})+\frac{2\mu}{\hbar^2}\left[E_{0\ell}-V_{eff}(\tilde{\varsigma})\right]=0
\label{E9}
\end{equation}
Since the logarithmic derivative $\phi_0(\tilde{\varsigma})$ for the ground state has one zero and no pole, it has to take the linear form in $\tilde{\varsigma}$. Thus, we assume the following solution for the ground states
\begin{equation}
\phi_0(\tilde{\varsigma})=A+B\tilde{\varsigma}
\label{E10}
\end{equation}
By putting equation (\ref{E10}) into (\ref{E9}), we can solve the non-linear Riccati equation. After proper comparison, it is straightforward to obtain the ground state energy and values of A and B as
\begin{equation}
E_{0\ell}=\mathcal{A}-\frac{\hbar^2A^2}{2\mu}\ \ \ \mbox{with}\ \ \ A=\frac{\mu}{\hbar^2}\frac{\mathcal{B}-\mathcal{F}/c_h^2}{B}+\frac{B}{2} \ \ \mbox{and}\ \ \ B=\frac{ac_h}{2r_e}+\frac{ac_h}{2r_e}\sqrt{1+\frac{8\mu \mathcal{F}r_e^2}{\alpha^2\hbar^2c_h^2}}.
\label{E11}
\end{equation}
Since we now have all basic ingredient required to perform our calculations, thus, we proceed to calculating integrals (\ref{E2})
\begin{eqnarray}
\int_{r_{a}}^{r_{b}}k(r)dr&=&-\frac{r_e}{\alpha}\int_{\tilde{\varsigma}_{a}}^{\tilde{\varsigma}_{b}}\frac{k(\tilde{\varsigma})}{a\tilde{\varsigma}(1+\tilde{\varsigma})}d\tilde{\varsigma}=-\frac{r_e}{\alpha}\int_{\tilde{\varsigma}_{0a}}^{\tilde{\varsigma}_{0b}}\sqrt{\frac{2\mu \mathcal{F}}{\hbar^2}}\frac{\left[-\left(\tilde{\varsigma}-\tilde{\varsigma}_a\right)\left(\tilde{\varsigma}-\tilde{\varsigma}_b\right)\right]^{1/2}}{\tilde{\varsigma}(1+\tilde{\varsigma}c_h)}d\tilde{\varsigma}\nonumber\\
&=&-\frac{\pi r_e}{\alpha}\sqrt{\frac{2\mu \mathcal{F}}{\hbar^2}}\left[\frac{\sqrt{(1+\tilde{\varsigma}_ac_h)(1+\tilde{\varsigma}_bc_h)}}{c_h}-\frac{1}{c_h}-\sqrt{\tilde{\varsigma}_a\tilde{\varsigma}_b}\right]\\
&=&-\frac{\pi r_e}{\alpha}\sqrt{\frac{2\mu \mathcal{F}}{\hbar^2}}\left[\frac{\sqrt{(\mathcal{A}-E_{n\ell})c_h^2-\mathcal{B}c_h+\mathcal{F}}}{\mathcal{F}c_h}-\frac{1}{c_h}-\sqrt{\frac{\mathcal{A}-E_{n\ell}}{\mathcal{F}}}\right],\nonumber
\label{E12}
\end{eqnarray}
where we have used the following standard integral
\begin{equation}
\int_{x_a}^{x_b}\frac{\sqrt{(x_a-x)(x-x_b)}}{x(1+Qx)}dx=\pi\left[\frac{\sqrt{(Qx_a+1)(Qx_b+1)}}{Q}-\frac{1}{Q}-\sqrt{x_ax_b}\right].
\label{E13}
\end{equation}
Furthermore, we can find the integral at the right hand side as 
\begin{eqnarray}
\int_{r_{0a}}^{r_{0a}}\phi(r)\left[\frac{dk_0(r)}{dr}\right]\left[\frac{d\phi(r)}{dr}\right]^{-1}dr&=&-\frac{r_e}{\alpha}\int_{\tilde{\varsigma}_{0a}}^{\tilde{\varsigma}_{0b}}\left[\frac{dk(r)}{dr}\right]\left[\frac{d\phi(\tilde{\varsigma})}{d\tilde{\varsigma}}\right]^{-1}\left(\frac{A}{B}+\tilde{\varsigma}\right)d\tilde{\varsigma}\nonumber\\
&=&\frac{r_e}{2\alpha}\int_{\tilde{\varsigma}_{0a}}^{\tilde{\varsigma}_{0b}}\sqrt{\frac{2\mu \mathcal{F}}{\hbar^2}}\frac{\left[2\tilde{\varsigma}-(\tilde{\varsigma}_a+\tilde{\varsigma}_b)\right]\left[\frac{A}{B}+\tilde{\varsigma}\right]}{\tilde{\varsigma}(1+\tilde{\varsigma}c_h)\sqrt{-(\tilde{\varsigma}-\tilde{\varsigma}_a)(\tilde{\varsigma}-\tilde{\varsigma}_b)}}d\tilde{\varsigma}\nonumber\\
&=&\frac{r_e}{2\alpha}\sqrt{\frac{2\mu \mathcal{F}}{\hbar^2}}\int_{\tilde{\varsigma}_{0a}}^{\tilde{\varsigma}_{0b}}\left[\left(\frac{A}{B}-\frac{1}{c_h}\right)\left(1+\frac{\tilde{\varsigma}_{0a}+\tilde{\varsigma}_{0b}}{2}c_h\right)\frac{1}{1+\tilde{\varsigma}c_h}+\frac{1}{c_h}\right.\nonumber\\
&&-\left.\frac{A}{B}\left(\frac{\tilde{\varsigma}_{0a}+\tilde{\varsigma}_{0b}}{2}\right)\frac{1}{\tilde{\varsigma}}\right]\times\frac{d\tilde{\varsigma}}{\sqrt{-(\tilde{\varsigma}-\tilde{\varsigma}_a)(\tilde{\varsigma}-\tilde{\varsigma}_b)}}\nonumber\\
&=&\left[\left(\frac{A}{B}-\frac{1}{c_h}\right)\left(1-\frac{\mathcal{B}c_h}{2\mathcal{F}}\right)\frac{1}{\sqrt{1-\frac{\mathcal{B}c_h}{\mathcal{F}}+\frac{\mathcal{A}-E_{0\ell}}{\mathcal{F}}c_h^2}}\right.\nonumber\\
&&\left.+\frac{A}{B}\left(\frac{\mathcal{B}}{2\mathcal{F}}\right)\frac{\sqrt{\mathcal{F}}}{{\sqrt{\mathcal{A}-E_{0\ell}}}{\mathcal{F}}}\right]\frac{\pi r_e}{\alpha}\sqrt{\frac{2\mu \mathcal{F}}{\hbar^2}}\nonumber\\
&=&\frac{\pi r_e}{\alpha}\sqrt{\frac{2\mu \mathcal{F}}{\hbar^2}}\left[\frac{1}{Bc_h}\sqrt{\frac{2\mu \mathcal{F}}{\hbar^2}}+\frac{1}{c_h}\right]
\label{E14}
\end{eqnarray}
Using the results in equations (\ref{E14}) and (\ref{E12}) with equation (\ref{E2}), we can find the energy energy eigenvalues equation for the sTW diatomic molecular potential as
\begin{eqnarray}
E_{n\ell}^{D}&=&\frac{\hbar^2\left(\tilde{\eta}^2-\frac{1}{4}\right)D_0}{2\mu r_e^2}-\frac{\alpha^2\hbar^2}{2\mu r_e^2}\left[\frac{(\delta+n)^2+\frac{\left(\tilde{\eta}^2-\frac{1}{4}\right)}{\alpha^2c_h^2}(D_1c_h-D_2)+\frac{2\mu V_er_e^2}{\alpha^2\hbar^2}\left(1-\frac{1}{c_h^2}\right)}{2(\delta+n)}\right]^2\label{E15}\\
&&\mbox{with}\ \ \delta=\frac{1}{2}+\frac{1}{2}\sqrt{1+\frac{4}{c_h^2}\left(\frac{D_2\left(\tilde{\eta}^2-\frac{1}{4}\right)}{\alpha^2}+\frac{2\mu V_er_e^2}{\alpha^2\hbar^2}(1-c_h)^2\right)}.\nonumber
\end{eqnarray}
In three-dimensions $(D = 3)$, it can be reduced to the form
\begin{eqnarray}
E_{n\ell}&=&\frac{\hbar^2\ell(\ell+1)D_0}{2\mu r_e^2}-\frac{\alpha^2\hbar^2}{2\mu r_e^2}\left[\frac{(\delta+n)^2+\frac{\ell(\ell+1)}{\alpha^2c_h^2}(D_1c_h-D_2)+\frac{2\mu V_er_e^2}{\alpha^2\hbar^2}\left(1-\frac{1}{c_h^2}\right)}{2(\delta+n)}\right]^2\label{E16}\\
&&\mbox{with}\ \ \delta=\frac{1}{2}+\frac{1}{2}\sqrt{1+\frac{4}{c_h^2}\left(\frac{D_2\ell(\ell+1)}{\alpha^2}+\frac{2\mu V_er_e^2}{\alpha^2\hbar^2}(1-c_h)^2\right)}.\nonumber
\end{eqnarray}
\section{The Eigenfunctions}
For the sake of completeness, we study the corresponding wavefunctions for this potential. For this purpose we introduce a new transformation of the form $t=e^{-b_h(r-r_e) }$ $\in(e^\alpha, 0)$ in equation (\ref{E7}), which maintained the finiteness of the transformed wave functions on the boundary conditions. Thus, we can find
\begin{eqnarray}
&&\frac{d^2U_{n\ell}(t)}{dt^2}+\frac{1}{t}\frac{dU_{n\ell}(t)}{dt}+\frac{1}{t^2(1-c_ht)^2}\left\{\left[\frac{2\mu r_e^2E_{n\ell}}{\hbar^2\alpha^2}-\frac{\left(\tilde{\eta}^2-\frac{1}{4}\right)}{\alpha^2}D_0\right]\right.\nonumber\\
&&\left.+\left[-2c_h\left(\frac{2\mu r_e^2(E_{n\ell}+V_e)}{\alpha^2\hbar^2}-\frac{\left(\tilde{\eta}^2-\frac{1}{4}\right)}{\alpha^2}D_0\right)+\frac{4\mu r_e^2V_e}{\hbar^2\alpha^2}-\frac{\left(\tilde{\eta}^2-\frac{1}{4}\right)}{\alpha^2}D_1\right]t\right.\label{E17}\\
&&\left.+\left[c_h^2\left(\frac{2\mu r_e^2(E_{n\ell}+V_e)}{\alpha^2\hbar^2}-\frac{\left(\tilde{\eta}^2-\frac{1}{4}\right)}{\alpha^2}D_0\right)+\frac{\left(\tilde{\eta}^2-\frac{1}{4}\right)}{\alpha^2}\left(D_1c_h-D_2\right)-\frac{2\mu r_e^2V_e}{\hbar^2\alpha^2}\right]t^2\right\}U_{n\ell}(t)=0.\nonumber
\end{eqnarray}
Following the procedure described in ref. \cite{BJ16}, we can write the solution $U_{n\ell}(t)$ in terms of hypergeometric polynomials and thus, the wave function takes the form
\begin{eqnarray}
U_{n\ell}(\zeta)=N_{n\ell}e^{-p\alpha \zeta}(1-c_he^{-\alpha \zeta})^q\ _2F_1\left(-n, n+2(p+q); 2p+1, c_he^{-\alpha \zeta}\right),
\label{E20}
\end{eqnarray}
with
\begin{equation}
p=\sqrt{\left[\left(\frac{\tilde{\eta}^2-\frac{1}{4}}{\alpha^2}\right)D_0-\frac{2\mu r_e^2}{\hbar^2\alpha^2}E_{n\ell}\right]}\ \ \mbox{and} \ \ q=\frac{1}{2}\left\{1+\sqrt{1+\frac{4}{c_h^2}\left[\left(\frac{\tilde{\eta}^2-\frac{1}{4}}{\alpha^2}\right)D_2+\frac{2\mu r_e^2V_e}{\hbar^2\alpha^2}(1-c_h)^2\right]}\right\},
\label{E21}
\end{equation}
where $N_{n\ell}$ is the normalization constant. For a further detail on the calculation of similar potential models solved using formula method, one is advised to refer to other work \cite{BJ16}
\begin{table}[!t]
{\scriptsize
\caption{Model parameters of the diatomic molecules studied in the present work. }\vspace*{10pt}
\begin{tabular}{cccccc}\hline\hline
{}&{}&{}&{}&{}&{}\\[-1.0ex]
Molecules(states)&$c_h$&$\mu/ 10^{-23}(g)$& $b_h(\AA^{-1})$&$r_e (\AA)$&$D(cm^{-1})$\\[2.5ex]\hline\hline
NO $\left(a^4\Pi_i\right)$	&0.0082003	&1.249	&2.408413	&1.451	&16361	\\[1ex]
NO $\left(B^2\Pi_r\right)$	&-0.482743	&1.249	&3.42650	&1.428	&22722\\[1ex]
NO $\left(L'^2\phi\right)$	&-0.073021	&1.249	&2.73796	&1.451	&14501\\[1ex]
NO $\left(b^4\Sigma^{-}\right)$	&-0.085078	&1.249	&3.01538	&1.318	&21183	\\[1ex]
ICl $\left(X^1\Sigma_g^{+}\right)$	&-0.086212	&4.55237	&2.008578	&2.3209	&17557\\[1ex]
ICl $\left(A^3\Pi_1\right)$	&-0.167208	&4.55237	&2.542557	&2.6850	&3814.7\\[1ex]
ICl $\left(A'^3\Pi_2\right)$	&-0.157361	&4.55237	&2.373450	&2.6650	&4875\\[1ex]
\hline\hline
\end{tabular}\label{tab1}
\vspace*{-1pt}}
\end{table}
\begin{table}[!t]
{\scriptsize
\caption{\normalsize The bound states energy eigenvalues ($D = 2$ and $3$) for  set of diatomic molecules for various $n$ and rotational $\ell$ quantum numbers in sDF diatomic molecular potential. } \vspace*{10pt}{
\begin{tabular}{cccccccccc}\hline\hline
{}&{}&{}&{}&{}&{}&{}&{}&{}&{}\\[-1.0ex]
D&$n$&$\ell$&NO $\left(a^4\Pi_i\right)$&NO $\left(B^2\Pi_r\right)$&NO $\left(L'^2\phi\right)$&NO $\left(b^4\Sigma^{-}\right)$&ICl $\left(X^1\Sigma_g^{+}\right)$&ICl $\left(A^3\Pi_1\right)$&ICl $\left(A'^3\Pi_2\right)$\\[2.5ex]\hline\hline
&	0	&0	&-1.971298585	&-2.88233770	&-1.855428285	&-2.695026855	&-2.200695845	&-0.4861418795	&-0.618432865\\[1ex]
&	1	&0	&-1.859308585	&-3.01386965	&-1.972859085	&-2.834614855	&-2.248819245	&-0.5129225795	&-0.646829015\\[1ex]
&		&1	&-1.859148585	&-3.01373250	&-1.972723585	&-2.834451355	&-2.248804845	&-0.5129118495	&-0.646818105\\[1ex]
&	2	&0	&-1.750608585	&-3.14759050	&-2.093631485	&-2.977436555	&-2.297420945	&-0.5403157895	&-0.675774025\\[1ex]
&		&1	&-1.750468585	&-3.14745280	&-2.093493685	&-2.977270755	&-2.297406745	&-0.5403049695	&-0.675763045\\[1ex]
&		&2	&-1.750098585	&-3.14703970	&-2.093081285	&-2.976773655	&-2.297363645	&-0.5402725095	&-0.675730195\\[1ex]
2&	3	&0	&-1.645178585	&-3.28347662	&-2.217724085	&-3.123472855	&-2.346500045	&-0.5683148895	&-0.705263165\\[1ex]
&		&1	&-1.645068585	&-3.28333841	&-2.217584185	&-3.123304955	&-2.346485745	&-0.5683039695	&-0.705252125\\[1ex]
&		&2	&-1.644678585	&-3.28292375	&-2.217165185	&-3.122801255	&-2.346442645	&-0.5682712195	&-0.705219005\\[1ex]
&		&3	&-1.644058585	&-3.28223274	&-2.216466785	&-3.121961855	&-2.346370345	&-0.5682166495	&-0.705163785\\[1ex]
&	4	&0	&-1.543078585	&-3.42150558	&-2.345115985	&-3.272704955	&-2.396055245	&-0.5969133695	&-0.735291785\\[1ex]
&		&1	&-1.542938585	&-3.42136686	&-2.344973985	&-3.272534755	&-2.396040945	&-0.5969023495	&-0.735280655\\[1ex]
&		&2	&-1.542598585	&-3.42095075	&-2.344548485	&-3.272024355	&-2.395997445	&-0.5968693195	&-0.735247275\\[1ex]
&		&3	&-1.541978585	&-3.42025729	&-2.343839285	&-3.271173455	&-2.395924945	&-0.5968142695	&-0.735191635\\[1ex]
&		&4	&-1.541118585	&-3.41928647	&-2.342846285	&-3.269982555	&-2.395823745	&-0.5967371995	&-0.735113745\\[1ex]
&	0	&0	&-1.971278585	&-2.88230356	&-1.855395085	&-2.694986655	&-2.200692145	&-0.4861392095	&-0.618430175\\[1ex]
&	1	&0	&-1.859278585	&-3.01383536	&-1.972825385	&-2.834574055	&-2.248815545	&-0.5129198795	&-0.646826285\\[1ex]
&		&1	&-1.859008585	&-3.01356106	&-1.972554585	&-2.834247155	&-2.248787045	&-0.5128984495	&-0.646804535\\[1ex]
&	2	&0	&-1.750568585	&-3.14755607	&-2.093596885	&-2.977395155	&-2.297417445	&-0.5403130995	&-0.675771295\\[1ex]
&		&1	&-1.750288585	&-3.14728068	&-2.093321985	&-2.977063955	&-2.297388745	&-0.5402914395	&-0.675749345\\[1ex]
&		&2	&-1.749808585	&-3.14672991	&-2.092771985	&-2.976400855	&-2.297331445	&-0.5402481695	&-0.675705525\\[1ex]
3&	3	&0	&-1.645158585	&-3.28344206	&-2.217689085	&-3.123430855	&-2.346496645	&-0.5683121595	&-0.705260395\\[1ex]
&		&1	&-1.644908585	&-3.28316563	&-2.217409585	&-3.123095155	&-2.346467645	&-0.5682903295	&-0.705238345\\[1ex]
&		&2	&-1.644418585	&-3.28261279	&-2.216850885	&-3.122423255	&-2.346410245	&-0.5682466595	&-0.705194145\\[1ex]
&		&3	&-1.643678585	&-3.28178358	&-2.216012785	&-3.121415755	&-2.346323545	&-0.5681811695	&-0.705127915\\[1ex]
&	4	&0	&-1.543028585	&-3.42147091	&-2.345080285	&-3.272662455	&-2.396051945	&-0.5969106095	&-0.735288995\\[1ex]
&		&1	&-1.542798585	&-3.42119348	&-2.344796885	&-3.272322355	&-2.396022645	&-0.5968885995	&-0.735266735\\[1ex]
&		&2	&-1.542328585	&-3.42063869	&-2.344229385	&-3.271641355	&-2.395964745	&-0.5968445495	&-0.735222225\\[1ex]
&		&3	&-1.541588585	&-3.41980654	&-2.343378185	&-3.270620455	&-2.395877945	&-0.5967784695	&-0.735155455\\[1ex]
&		&4	&-1.540608585	&-3.41869709	&-2.342243285	&-3.269259655	&-2.395761945	&-0.5966903895	&-0.735066435\\[1ex]
\hline\hline
\end{tabular}\label{tab2}}
\vspace*{-1pt}}
\end{table}
\begin{table}[!t]
{\scriptsize
\caption{\normalsize The bound states energy eigenvalues $D = 4$ and $5$) for set of diatomic molecules for various $n$ and rotational $\ell$ quantum numbers in sDF diatomic molecular potential. } \vspace*{10pt}{
\begin{tabular}{cccccccccc}\hline\hline
{}&{}&{}&{}&{}&{}&{}&{}&{}&{}\\[-1.0ex]
D&$n$&$\ell$&NO $\left(a^4\Pi_i\right)$&NO $\left(B^2\Pi_r\right)$&NO $\left(L'^2\phi\right)$&NO $\left(b^4\Sigma^{-}\right)$&ICl $\left(X^1\Sigma_g^{+}\right)$&ICl $\left(A^3\Pi_1\right)$&ICl $\left(A'^3\Pi_2\right)$\\[2.5ex]\hline\hline
&	0	&0	&-1.971158585	&-2.88220112	&-1.855295085	&-2.694865755	&-2.200681645	&-0.4861312495	&-0.618422105\\[1ex]
&	1	&0	&-1.859148585	&-3.01373250	&-1.972723585	&-2.834451355	&-2.248804845	&-0.5129118495	&-0.646818105\\[1ex]
&		&1	&-1.858778585	&-3.01332103	&-1.972317885	&-2.833961155	&-2.248762245	&-0.5128796695	&-0.646785525\\[1ex]
&	2	&0	&-1.750468585	&-3.14745280	&-2.093493685	&-2.977270755	&-2.297406745	&-0.5403049695	&-0.675763045\\[1ex]
&		&1	&-1.750098585	&-3.14703970	&-2.093081285	&-2.976773655	&-2.297363645	&-0.5402725095	&-0.675730195\\[1ex]
&		&2	&-1.749438585	&-3.14635128	&-2.092393885	&-2.975945655	&-2.297292045	&-0.5402183995	&-0.675675405\\[1ex]
3&	3	&0	&-1.645068585	&-3.28333841	&-2.217584185	&-3.123304955	&-2.346485745	&-0.5683039695	&-0.705252125\\[1ex]
&		&1	&-1.644678585	&-3.28292375	&-2.217165185	&-3.122801255	&-2.346442645	&-0.5682712195	&-0.705219005\\[1ex]
&		&2	&-1.644058585	&-3.28223274	&-2.216466785	&-3.121961855	&-2.346370345	&-0.5682166495	&-0.705163785\\[1ex]
&		&3	&-1.643188585	&-3.28126535	&-2.215488785	&-3.120786155	&-2.346269545	&-0.5681402395	&-0.705086495\\[1ex]
&	4	&0	&-1.542938585	&-3.42136686	&-2.344973985	&-3.272534755	&-2.396040945	&-0.5969023495	&-0.735280655\\[1ex]
&		&1	&-1.542598585	&-3.42095075	&-2.344548485	&-3.272024355	&-2.395997445	&-0.5968693195	&-0.735247275\\[1ex]
&		&2	&-1.541978585	&-3.42025729	&-2.343839285	&-3.271173455	&-2.395924945	&-0.5968142695	&-0.735191635\\[1ex]
&		&3	&-1.541118585	&-3.41928647	&-2.342846285	&-3.269982555	&-2.395823745	&-0.5967371995	&-0.735113745\\[1ex]
&		&4	&-1.540048585	&-3.41803836	&-2.341569785	&-3.268451355	&-2.395693345	&-0.5966380895	&-0.735013565\\[1ex]
&	0	&0	&-1.970998585	&-2.88203041	&-1.855128785	&-2.694664455	&-2.200663845	&-0.4861179795	&-0.618408615\\[1ex]
&	1	&0	&-1.859008585	&-3.01356106	&-1.972554585	&-2.834247155	&-2.248787045	&-0.5128984495	&-0.646804535\\[1ex]
&		&1	&-1.858508585	&-3.01301246	&-1.972013385	&-2.833593555	&-2.248729945	&-0.5128555495	&-0.646761065\\[1ex]
&	2	&0	&-1.750288585	&-3.14728068	&-2.093321985	&-2.977063955	&-2.297388745	&-0.5402914395	&-0.675749345\\[1ex]
&		&1	&-1.749808585	&-3.14672991	&-2.092771985	&-2.976400855	&-2.297331445	&-0.5402481695	&-0.675705525\\[1ex]
&		&2	&-1.749048585	&-3.14590380	&-2.091946885	&-2.975407055	&-2.297245545	&-0.5401832395	&-0.675639815\\[1ex]
4&	3	&0	&-1.644908585	&-3.28316563	&-2.217409585	&-3.123095155	&-2.346467645	&-0.5682903295	&-0.705238345\\[1ex]
&		&1	&-1.644418585	&-3.28261279	&-2.216850885	&-3.122423255	&-2.346410245	&-0.5682466595	&-0.705194145\\[1ex]
&		&2	&-1.643678585	&-3.28178358	&-2.216012785	&-3.121415755	&-2.346323545	&-0.5681811695	&-0.705127915\\[1ex]
&		&3	&-1.642668585	&-3.28067802	&-2.214895385	&-3.120072855	&-2.346208245	&-0.5680938395	&-0.705039565\\[1ex]
&	4	&0	&-1.542798585	&-3.42119348	&-2.344796885	&-3.272322355	&-2.396022645	&-0.5968885995	&-0.735266735\\[1ex]
&		&1	&-1.542328585	&-3.42063869	&-2.344229385	&-3.271641355	&-2.395964745	&-0.5968445495	&-0.735222225\\[1ex]
&		&2	&-1.541588585	&-3.41980654	&-2.343378185	&-3.270620455	&-2.395877945	&-0.5967784695	&-0.735155455\\[1ex]
&		&3	&-1.540608585	&-3.41869709	&-2.342243285	&-3.269259655	&-2.395761945	&-0.5966903895	&-0.735066435\\[1ex]
&		&4	&-1.539388585	&-3.41731030	&-2.340824985	&-3.267557955	&-2.395617345	&-0.5965802895	&-0.734955135\\[1ex]\hline\hline
\end{tabular}\label{tab3}}
\vspace*{-1pt}}
\end{table}
\section{Results and Conclusions}
In this study, in an attempt to find a more suitable potential that stimulate the atomic interaction in diatomic molecules, we suggested sTW diatomic molecular potential as a modification for the TW diatomic molecular potential. The bound state solution of this potential has been found in an arbitrary D-dimension via the improved exact quantization rule.  

Further, using the spectroscopic parameters presented in table \ref{tab1} which are taken from ref. \cite{J14}, we computed rotational-vibrational energy spectrum of some diatomic molecules in 2,3,4,5-dimensions. The results are presented in tables \ref{tab2} and \ref{tab3}. In our numerical computations, we have used the following conversions: 1amu $= 931.494 028 MeV/c^2$, $1cm^{-1} = 1.239841875\times10^{-4}$eV, and $\hbar c = 1973.29eV\AA$.

From equation (\ref{E15}), it can be seen that two interdimensional states are degenerate whenever $(n,\ell,D)\rightarrow(n,\ell\pm1,D\mp2)\Rightarrow E_{n,\ell}^D=E_{n,\ell\pm1}^{(D\mp2)}$. Thus, a knowledge of $E_{n,\ell}^D$ for $D = 2$ to $5$ provides the information necessary to find $E_{n,\ell}^D$ for other higher dimensions.
For example, $E^{(2)}_{0,4} = E^{(4)}_{0,3} = E^{(6)}_{0,2} = E^{(8)}_{0,1}$ . This is the same transformational invariance described for bound states of free atoms and molecules \cite{BJ22, BJ23, BJ24} and demonstrates the existence of interdimensional degeneracies among states of the confined Hulth$\acute{e}$n potential.

The advantage of the approach employed in this study is that it gives the eigenvalues through the calculation of two integral given by equation (\ref{E2}) and solving the resulting algebraic equation. Firstly, we can easily obtain the quantum correction by only considering the solution of the ground state of the quantum system since it is independent of the number of nodes of the wave function for exactly solvable quantum system. The general expressions obtained for the energy eigenvalues and wave functions can be easily reduced to the 3D space (D = 3) and for s-wave (i.e. $\ell=0$ state). The EQR produce as good results as the PQR, however the procedure followed using PQR is more shorter and quick.

\section*{Acknowledgments}
We thank the kind referees for the positive enlightening comments and suggestions, which have greatly helped us in making improvements to this paper. In addition, BJF acknowledges eJDS (ICTP).

\end{document}